\pdfoutput=1
\documentclass[aps,prl,amsmath,amssymb,reprint,groupedaddress]{revtex4-1}

\usepackage{xcolor}
\definecolor{blue(pigment)}{rgb}{0.2, 0.2, 0.6}
\usepackage[utf8]{inputenc}
\usepackage{graphicx}
\usepackage{dcolumn}
\usepackage{bm}
\usepackage[breaklinks=false,colorlinks=true,linkcolor=blue(pigment),urlcolor=blue(pigment),citecolor=blue(pigment)]{hyperref}

\begin{document}

\preprint{}

\title{Dynamics of trapped one-dimensional bosons for intermediate repulsive interactions}


\author{D.~Efe~Gökmen}
\email{dgoekmen@student.ethz.ch}
\affiliation{Department of Physics, ETH Zürich Hönggerberg, CH-8093 Zurich, Switzerland}
\author{M.~Cemal~Yalabik}
\affiliation{Department of Physics, Bilkent University, 06810 Ankara, Turkey}

\date{\today}

\begin{abstract}
Time-evolution of a few number of interacting, harmonically confined one-dimensional bosons is numerically obtained for arbitrary two-body $\delta-$potential interaction strengths. It is demonstrated that the period of the motion in a Newton's cradle configuration undergoes two crossovers as a function of interactions. Furthermore, through the evaluation of the structure factor, the dependence of Bragg-scattering peaks on the interaction strength ranging from the weak coupling regime to the impenetrable Tonks-Girardeau case is illustrated.
\end{abstract}

\pacs{}

\maketitle




Experiments on one-dimensional (1D) bose gases \cite{1d_bec} demonstrated peculiar properties that are missing in their higher dimensional counterparts. In the absence of an external trap, the 1D bose gas with two-body contact interactions of any strength is described by the exactly integrable Lieb-Liniger (LL) model \cite{LL1, *LL2}.  Through the work of Olshanii \cite{atomic_scattering}, it has became possible to relate the interaction strength parameter $c$ of LL model to the real experimental parameters of ultracold gases in 1D traps. In practice, $c$ can be tuned via Feshbach resonances \cite{feshbach}. Since then, the infinitely strong repulsive interaction limit of the trapped LL gas, also known as the Tonks-Girardeau (TG) regime \cite{trapped-tg, *fermi-bose-review}, has been experimentally realized in optical traps \cite{tg_realisation,kinoshita_newtoncradle,*Kinoshita1125}. Recently, an experimental implementation of the quantum Newton's cradle of tunable interaction strength has been achieved using the dipole-diple interaction between highly magnetic dysprosium atoms \cite{PhysRevX.8.021030}. 

\begin{figure}
	\centering
	\includegraphics[width=1\linewidth]{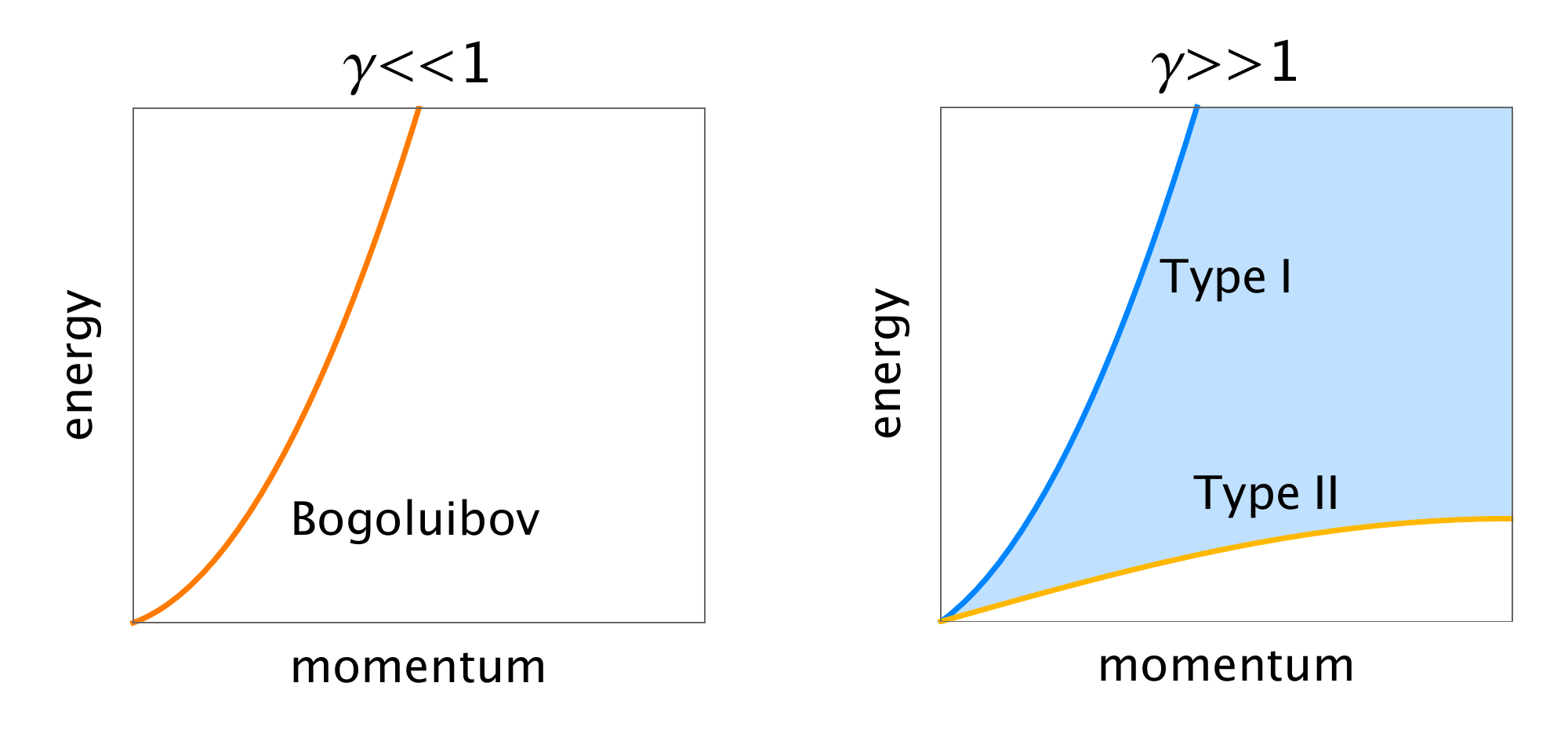}
	\caption{(Color online) The dispersion of two types of excitation modes in the Lieb-Liniger model. In the weakly interacting ($\gamma\ll1$) regime, the excitation is in good agreement with the Bogoliubov approximation. For strong interaction ($\gamma\gg1$), there is a continuum of modes enveloped by Type I quasiparticle and Type II quasihole excitations.}
	\label{fig:ll}
\end{figure}

In spite of the rich history backing the subject \cite{RevModPhys.83.1405}, most of the studies on the thermodynamic and spectral properties of interacting 1D bose gases have focused on the limiting cases of impenetrable TG regime \cite{finite-T-tg,PhysRevA.96.013613,bragg-tg,PhysRevA.91.063619} and the weakly interacting regime which is well described by Bogoliubov theory. As shown in Fig. \ref{fig:ll}, the strongly interacting regime broadens the Bogoliubov spectrum into a continuum between two types of modes. Type I modes are bosonic quasiparticle modes and the Type II modes are fermionic quasihole modes \cite{LL2}.

As far as the present authors can ascertain, it has not yet been resolved whether the finite mutual interaction strength compromises the integrability of a system of bosons in the presence of an external potential. On the other hand, numerical approaches proved powerful in shedding light into the crossover between weak interaction and TG regimes \cite{PhysRevA.74.053612,*PhysRevLett.100.040401}. In particular, the intermediate window of momentum distributions has only been investigated numerically \cite{PhysRevA.68.031602,PhysRevA.75.021601}, despite not being able to describe all physics from infrared to the ultraviolet range. In this Letter, three main results are presented: an analysis of the real-time dynamics reduced density matrix of a trapped 1D few-boson system, the dependence of the motion period on the interaction strength $c$, and the full range structure factor at finite $c$.

\begin{figure*}
	(a)
	\begin{minipage}[b]{0.33\linewidth}
		\includegraphics[width=\linewidth]{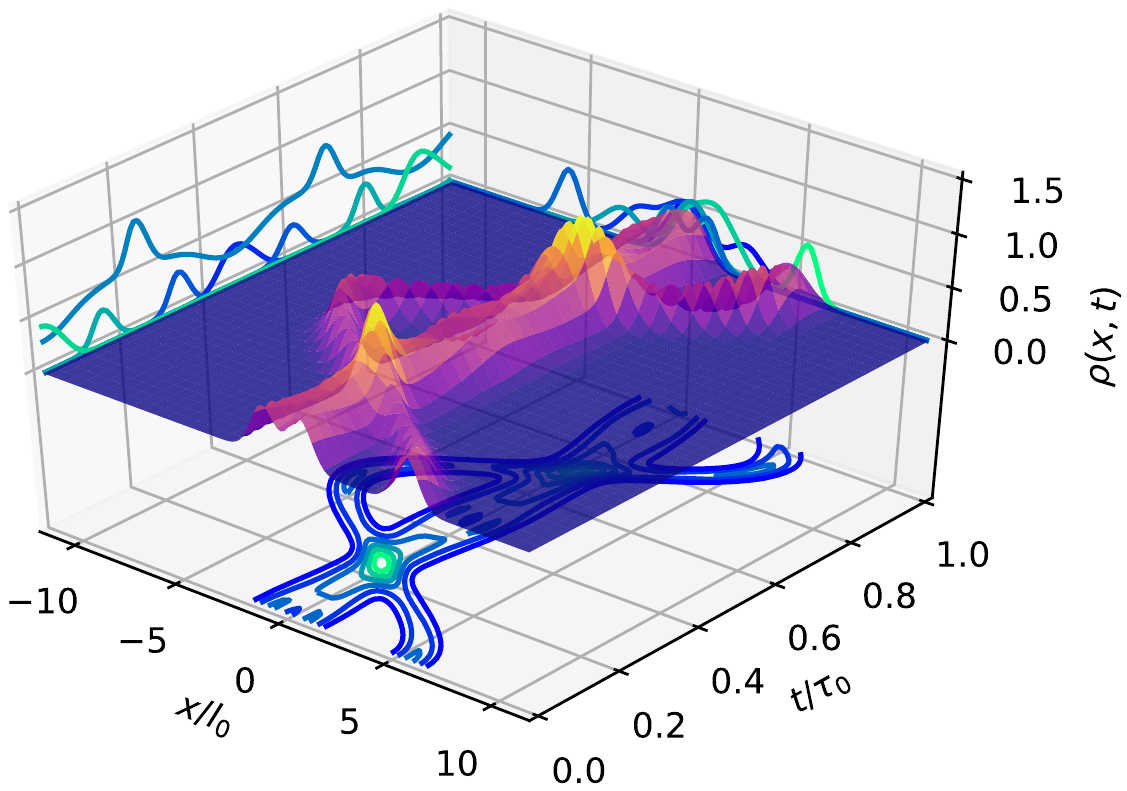}
	\end{minipage}
	\begin{minipage}[b]{0.33\linewidth}
		\includegraphics[width=\linewidth]{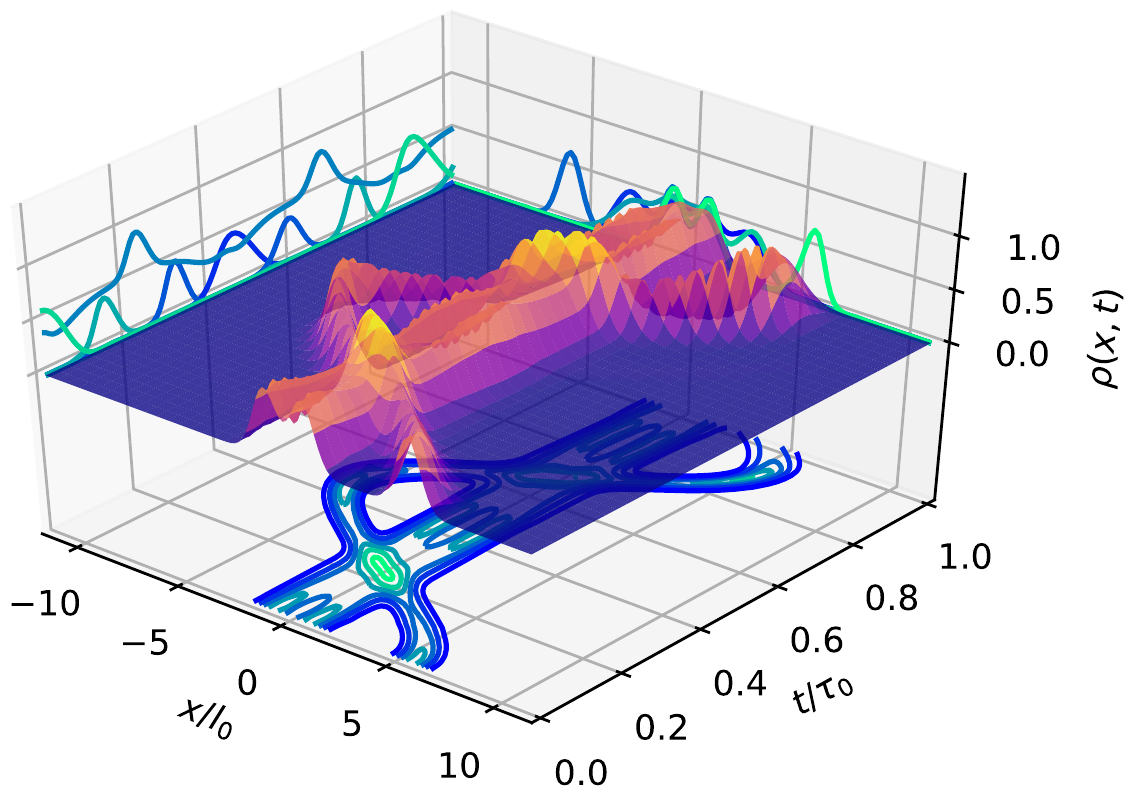}
	\end{minipage}
	(b)
	\begin{minipage}[b]{0.26\linewidth}
		\includegraphics[width=\linewidth]{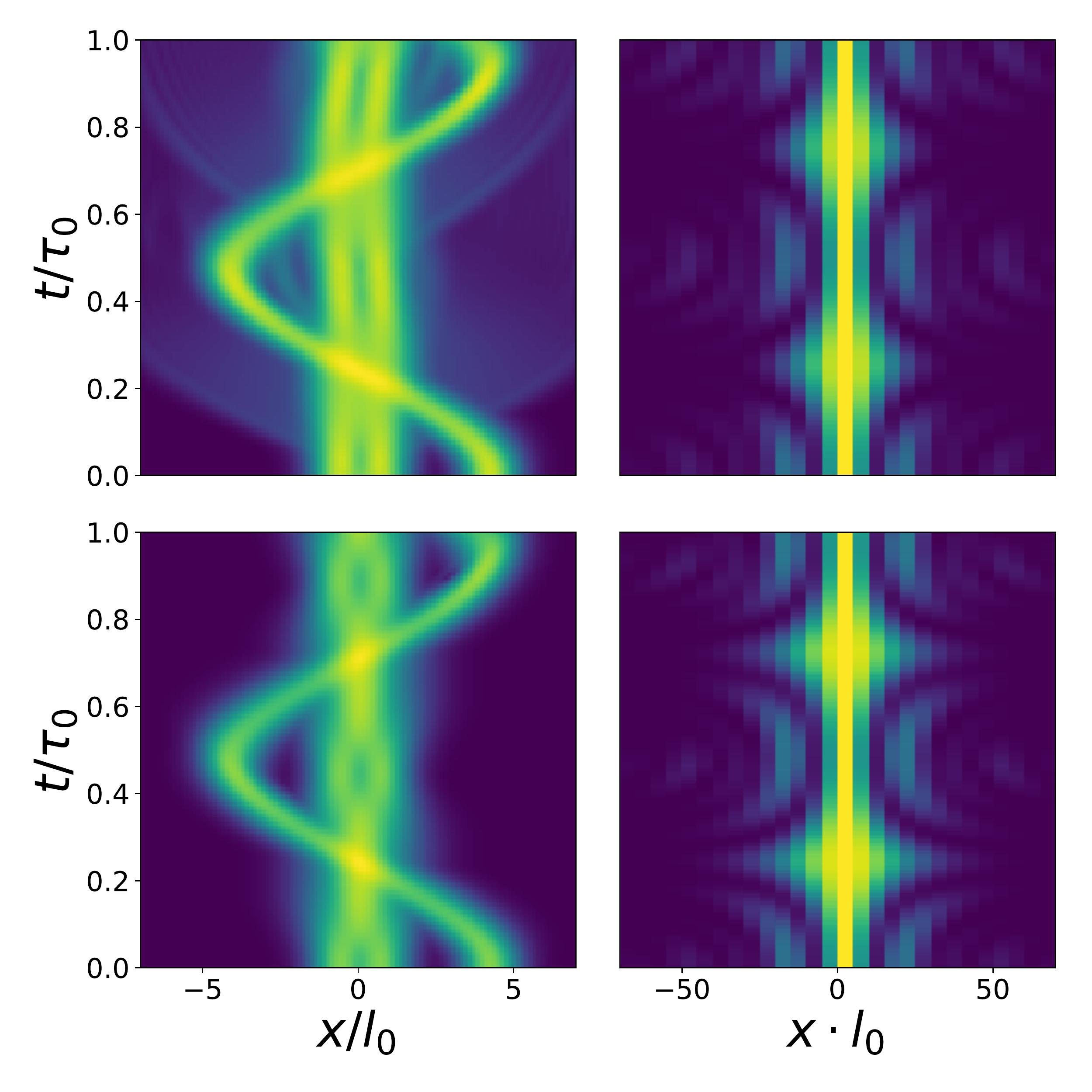}
	\end{minipage}	
	\caption{\label{realspace} (Color online) (a) Time evolution of the spatial density profile of the three-boson system initially in a Newton's cradle configuration for (left) $\gamma=0.3$ and (right) $\gamma=3$. (b) Dynamics of the bosonic Newton's cradle at interaction strengths $\gamma=0.003$ (upper panel) and $\gamma=1000$ (lower panel). The left panel is the evolution of the real-space density profile $\rho(x,t)$. The right panel is the evolution of the structure factor $F(\Delta k,t)$ which is related to the rate of momentum transfer upon Bragg reflections.}
\end{figure*}

\paragraph{Model and the numerical method.} In 1D, $N$ indistinguishable bosons of mass $m$ with repulsive two-body contact interaction $c\geq0$ inside a harmonic trap of characteristic frequency $\omega_0$ can be modelled by the Hamiltonian
\begin{equation}\label{hamiltonian}
\mathcal{H}(\mathbf{r})=\sum^N_{i=1}\left[-\frac{\hbar^2}{2m}\frac{\partial^2}{\partial x_i^2}+\frac{1}{2}m\omega_0^2x_i^2\right] + c\sum^N_{\scriptscriptstyle{i\leq j}} \delta(x_i-x_j),
\end{equation}
where $\mathbf{r}=(x,x_2,\cdots, x_N)$. It is convenient to define a unitless interaction parameter $\gamma=2c\left(\frac{\hbar^2}{2m \Delta x^2}\right)^{-1}$, where $\Delta x$ is the spatial width that is used in numerical discretization. In the numerical implementation, the $\delta-$potential is alleviated by using a more realistic normalized interaction with a small, but finite range. Normalization ensures that in the zero-width limit it approaches to the $\delta-$distribution.

The symmetrized initial state $\Psi(\mathbf{r},t=0)$ is composed of the Slater permanent of $N$ localized wavepackets $\psi_j$, with $j\in [1,N]$. In particular, the Newton's cradle configuration comprises one \textit{swinging} wavepacket displaced by $\alpha$ with respect to the equilibrium position
\begin{equation}\label{init1}
\psi_1(x_1)=A\exp{\left[-\frac{(x_1+\alpha)^2}{2\sigma^2}\right]},
\end{equation}
corresponding to the lifted ball in the cradle. The remaining $N-1$ wavepackets are nearly stationary and centered around the equilibrium, each separated by small distance $\epsilon \ll \alpha$
\begin{equation}\label{init2}
\psi_i(x_i)=A\exp{\left[-\frac{[x_i+2(N/2-i)\epsilon]^2}{2\sigma^2}\right]},
\end{equation}
with $i\in [2,N]$. Here, $A$ is the normalization, and $\sigma$ is the wavepacket width. Moreover it is taken that $\epsilon/\sigma \approx 1$, because $\epsilon$ is related to an effective radius of the bosons which tends to $\sigma$ as $\gamma\rightarrow\infty$. The goal is to study the dynamics of this system for different values of the interaction parameter $\gamma$ by calculating the time evolution of $\Psi(\mathbf{r},t)$ as governed by the time dependent Schrödinger equation (TDSE).

Diffusion Monte Carlo techniques \cite{PhysRevA.89.063616,PhysRevA.9.2178,PhysRevA.68.031602,PhysRevA.92.021601} and density matrix renormalization group \cite{PhysRevA.75.021601} are common methods to extract the ground state and low-energy excitations in interacting 1D bosonic systems. In addition, the few-body case with a double-well trap has been investigated using the Multi-Configuration Time-Dependent Hartree \cite{MEYER199073} method by Zöllner {\em et al.} \cite{PhysRevA.74.053612,*PhysRevLett.100.040401}. In this study, on the other hand, the $N$-particle TDSE $\mathcal{H}(\mathbf{r},t)\Psi(\mathbf{r},t)=i\hbar\frac{\partial}{\partial t}\Psi(\mathbf{r},t)$ is numerically integrated for low $N$ via a split operator method. Here, the time evolution according to $\mathcal{H}$ of Eq. \ref{hamiltonian} is obtained by alternated advancement in real and Fourier spaces for $\Psi(\mathbf{r},t=0)=\sum_{\mathcal{P}}\prod_{j=1}^N\,\psi_j(x_i)$ via Equations \ref{init1}, \ref{init2}. Here, $\mathcal{P}$ denotes the particle permutations. The premise of this method is using the fast Fourier transform to diagonalize the propagator $U(\Delta t)$ by breaking it up into the kinetic ($\mathcal{H}_{\rm k}$) and the potential ($\mathcal{H}_{\rm p}$) parts using Baker-Campbell-Hausdorff identity
\begin{equation}
\begin{split}
	U(\Delta t)& =  e^{-\frac{i}{\hbar}(\mathcal{H}_{\rm k}+\mathcal{H}_{\rm p})\Delta t}\\
	& \simeq U_{\rm TS}(\Delta t) \equiv e^{-\frac{i}{2\hbar}\mathcal{H}_{\rm p}\Delta t}e^{-\frac{i}{\hbar}\mathcal{H}_{\rm k}\Delta t}e^{-\frac{i}{2\hbar}\mathcal{H}_{\rm p}\Delta t}.
\end{split}
\end{equation}
The approximate expression $U_{\rm TS}(\Delta t)$ is also known as the Trotter-Suzuki expansion \cite{trotter1959product, *suzuki1976generalized}. The wavefunction is iteratively advanced in time by $\Psi(\mathbf{r},\Delta t)=U_{\rm TS}(\Delta t)\Psi(\mathbf{r},t=0)$ at the cost of an error proportional to $\Delta t^3$. Here, it is chosen that $\Delta t= t/n$ for the total number $n$ of time steps of the simulation. The expansion approaches the exact expression as $n\rightarrow\infty$ due to the Lie product formula.

The $N$-particle wave-function $\Psi(\mathbf{r},t)$ is computed over several cycles of motion for specific values of $\gamma$. Subsequently, the reduced real-space density matrix $\rho(x,x',t)$ is given by
\begin{equation}\label{realdensity}
\begin{split}
	&\rho(x,x',t)\\
	&=\int \Psi^*(x,\cdots, x_N,t)\Psi(x',\cdots, x_N,t) dx_2 \cdots dx_{N} .
\end{split}
\end{equation}
In what follows, the case where $x=x'$ will be considered, such that $\rho(x,x,t):=\rho(x,t)$ which describes the local density profile of the system.

\paragraph{Time evolution of density profiles.} The simulated space-time profile of $\rho(x,t)$ for $N=3$ is plotted in Fig. \ref{realspace}, and in the left panel of Fig. \ref{realspace}.b for $\gamma=0.3$ and for $\gamma=3$, which is one of the main results of this Letter. Here, the time and the space axes are respectively scaled in terms of the motion period $\tau_0=2\pi/\omega_0$ and the oscillator length $l_0=\sqrt{\hbar/m \omega_0}$, both of which calculated for $\gamma=0$ case. It is seen that in the limit $\gamma\to \infty$, for the aformentioned inital conditions, $\rho(x,t)$ assumes a periodic profile reminiscent of the worldlines of a classical Newton's cradle. On the other hand, tuning $\gamma$ results in alterations in the general form of the real-space density, as well as the motion period.

\begin{figure}
	(a)
	\begin{minipage}[t]{0.27\linewidth}
		\includegraphics[width=\linewidth]{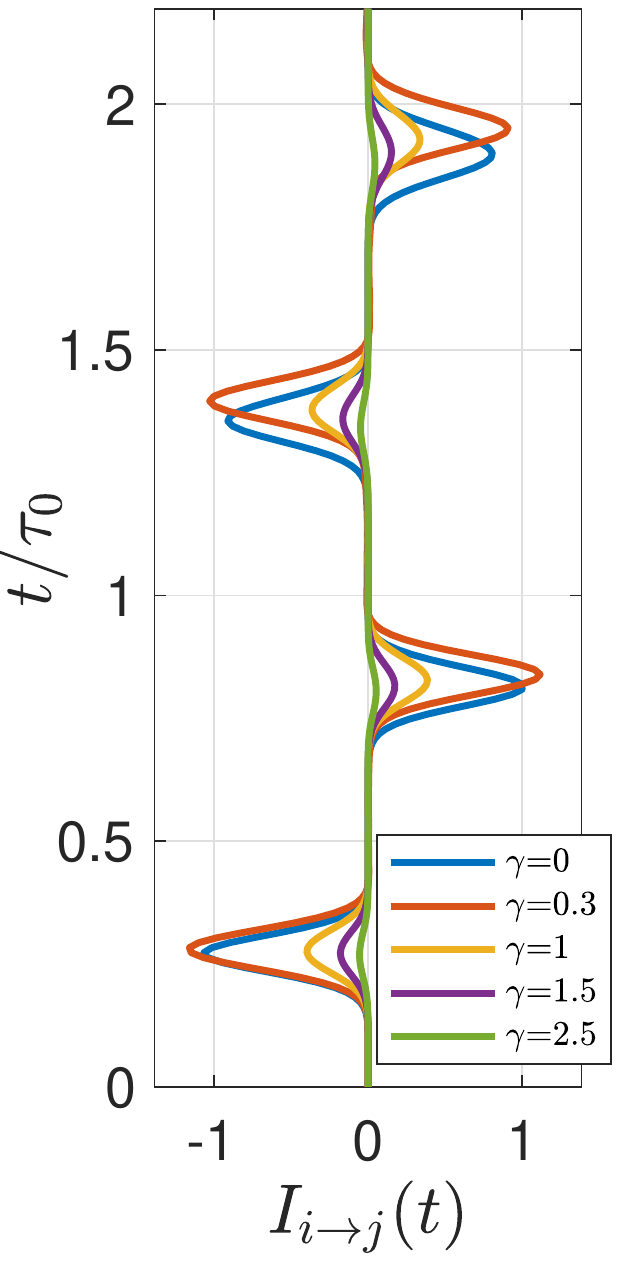}
	\end{minipage}
	(b)
	\hspace{-2mm}
	\begin{minipage}[b]{0.61\linewidth}
		\includegraphics[width=\linewidth]{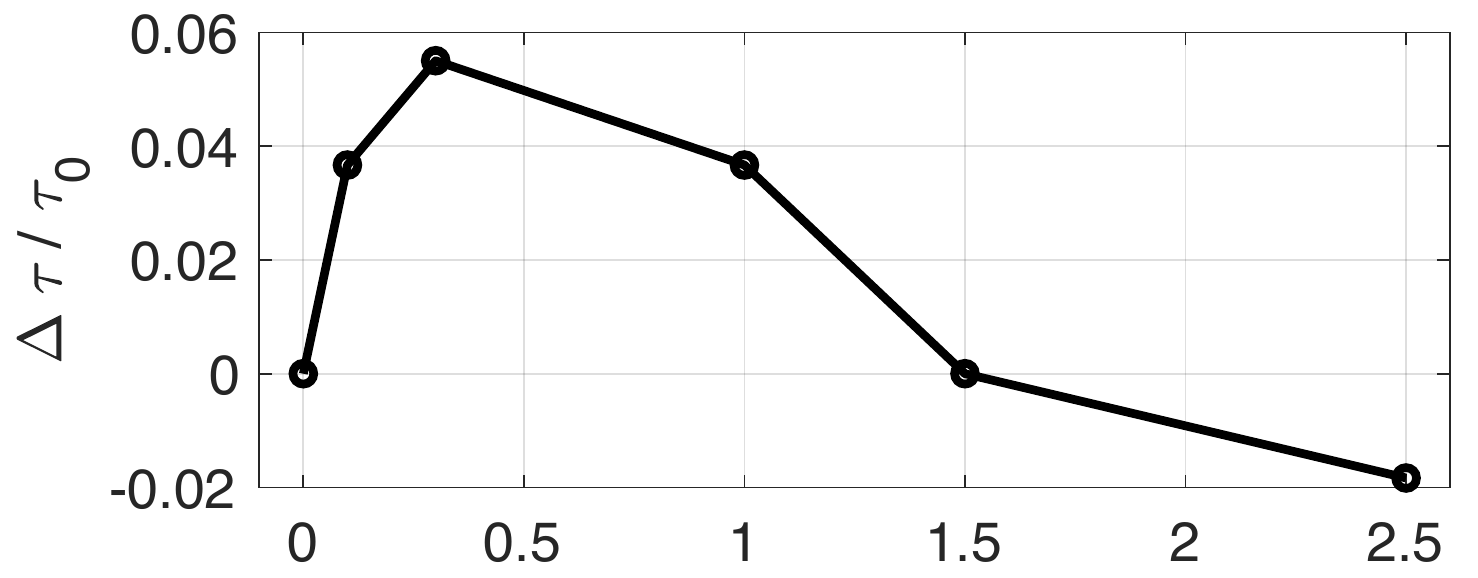}
		\includegraphics[width=\linewidth]{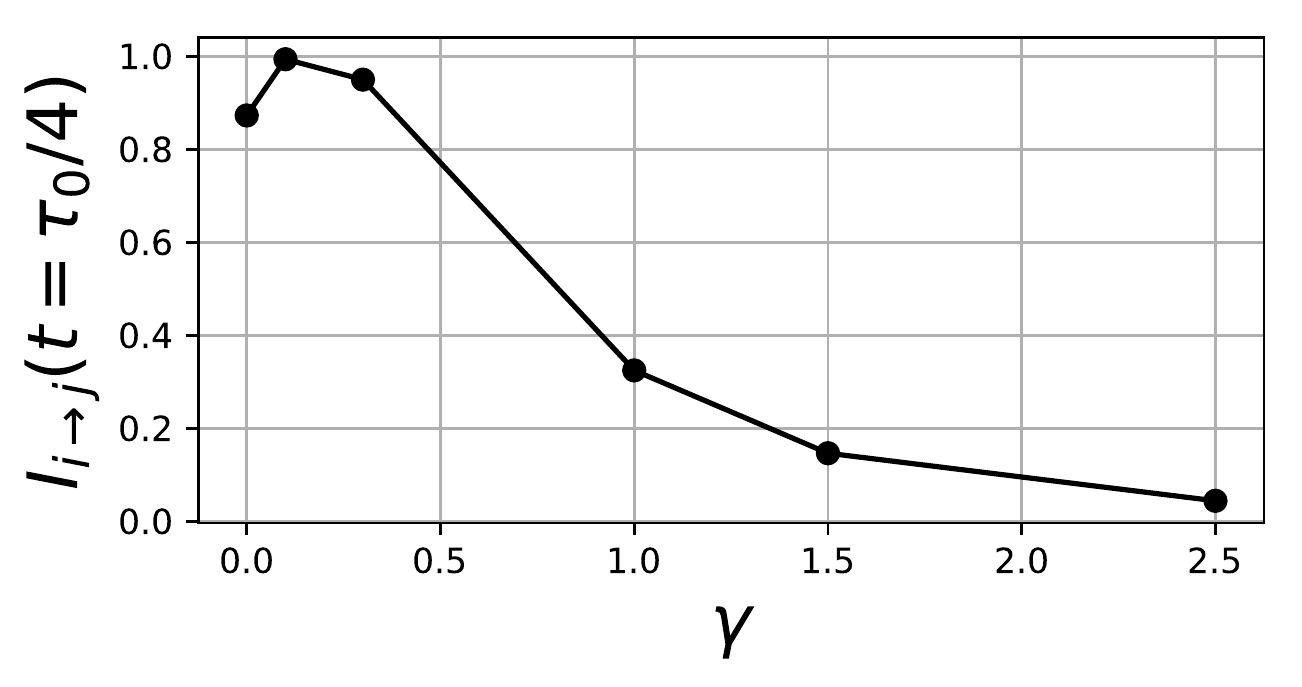}
	\end{minipage}
	\caption{\label{main} (Color online) $N=3$. (a) The plot of the probability current $I_{i\rightarrow j}(t)$ at $x_i=x_j$. It is understood that the likelihood of transmission of the $i$th boson through the $j$th boson decreases with stronger interaction. (b) (Upper panel) The deviation of the motion period $\tau$ from the one of non-interacting case $\tau_0$ as a function of the interaction strength $\gamma$. (Lower panel) The dependence of the maximal transmission amplitude on $\gamma$.}
\end{figure}

First observation is a qualitative one; the interference between the spacetime densities of the particles reduces with increasing $\gamma$. This can be seen for $N=3$ in the left panel of Fig. \ref{realspace}. It is seen that for $\gamma=3$ the real-space densities of the two central bosons cease to overlap. The depletion of the overlap region leads bosons to isolate each other, as $N=3$ humps emerge in $\rho(x,t)$. This process, referred to as \textit{fragmentation}, also signals that bosons are more likely to get reflected upon collisions. In fact, the $\gamma\to \infty$ limit imitates the Pauli exclusion principle, as fragmentation saturates into a so-called \textit{fermionized} state \cite{fermi-bose-tg}. On the other hand, for $\gamma=0.3$, such an overlap occurs at each event of collision, thereby the $N=3$ humps in $\rho(x,t)$ merge into a single one (see Fig. \ref{realspace}), indicating a finite transmission amplitude.

In addition to fragmentation and the alterations in transmission rates, the motion period deviates from $\tau_0$ as a function of $\gamma$ as shown in the lower panel of Fig. \ref{main}.b. This is the second main result of the present work. Here, two crossovers for the frequency deviations between zero to weak interaction, and weak to strong interaction are demonstrated. The results indicate that a maximal deviation at an intermediate value $\gamma \approx 0.4$ is followed by a return towards the initial value. Moreover, a minor reduction of $\tau$ at strong interactions signals a scattering resonance. The greatest deviation from $\tau_0$ is determined to be 5.8\%. Note that $\tau_0$ is equivalent to the oscillation period of a single particle, which is readily attained from the trap parameters in experiments. The measurement of deviations should reveal the empirical connection of the numerical parameter $\gamma$. The present results are in accordance with the previous experimental \cite{PhysRevLett.91.250402, Haller1224,PhysRevLett.113.035301} results along with the numerical \cite{PhysRevA.88.043601} and theoretical \cite{PhysRevA.66.043610,PhysRevA.92.021601} studies focusing on either the lower or the higher crossover for breathing modes.

\paragraph{Tunneling rates.} The time dependent amplitude profiles given in Fig. \ref{realspace} by themselves do not yield quantitative information about the transmission rates of particles at a collision event, which occurs when their wavepackets overlap. To obtain this quantity, a convenient starting point is to calculate the rate of change of the probability of finding a particle in state $\Psi(\mathbf{r},t)$, {\em i.e.} the probability current. From the continuity equation, the \textit{total} probability current density $\mathbf{J}(\mathbf{r},t)$ can be found to be
\begin{equation}
\mathbf{J}(\mathbf{r},t)=\sum_{i=1}^{N}\hat{\mathbf{x}}_iJ_i(\mathbf{r},t)=\frac{\hbar}{2m}\Im\{\Psi^*(\mathbf{r},t)\nabla\Psi(\mathbf{r},t)\},
\end{equation}
where $\hat{\mathbf{x}}_i$ is the unit vector along the coordinate $x_i$, and $\Im$ denotes the imaginary part. The components of $\mathbf{J}$ can be associated with each of the $N$ bosons. Therefore, the \textit{partial} probability density $J_{i}(x,\cdots,x_i,x_j=x_i\cdots,x_N,t)$ can be interpreted as a measure of the probability current density of $i$th particle \textit{through} $j$th particle upon their head on collision. In other words, by calculating the probability current $I_{i\rightarrow j}(t)$, which is the total flux of $\mathbf{J}$ through the interaction plane $x_i=x_j$ given by
\begin{equation}
\begin{split}
I_{i\rightarrow j}(t)= & \frac{\hbar}{2m}\int dx \cdots dx_{j-1} dx_{j+1} \cdots dx_N \\  & \times \Im \left\{\Psi^*(\mathbf{r},t)\frac{\partial}{\partial x_i}\Psi(\mathbf{r},t)\right\}_{x_i=x_j},
\end{split}
\end{equation}
one can quantify the likelihood of transmission of one boson through another at a collision. Simulated results for $I_{i\rightarrow j}(t)$ corresponding to different values of $\gamma$ are presented in Fig. \ref{main}.a. It is found that the transmission probability $\mathcal{T}(t)\propto|I_{i\rightarrow j}(t)|^2$ decreases with increasing $\gamma$ and saturates at zero for $\gamma \gtrsim 3$, corresponding to impenetrable bosons. Being in a low energy scattering regime, this limit, where $\mathcal{T}(t)$ approaches to zero, corresponds to the TG gas, as suggested by Olshanii\cite{atomic_scattering}.\\

\begin{figure}[]
	\includegraphics[width=1\linewidth]{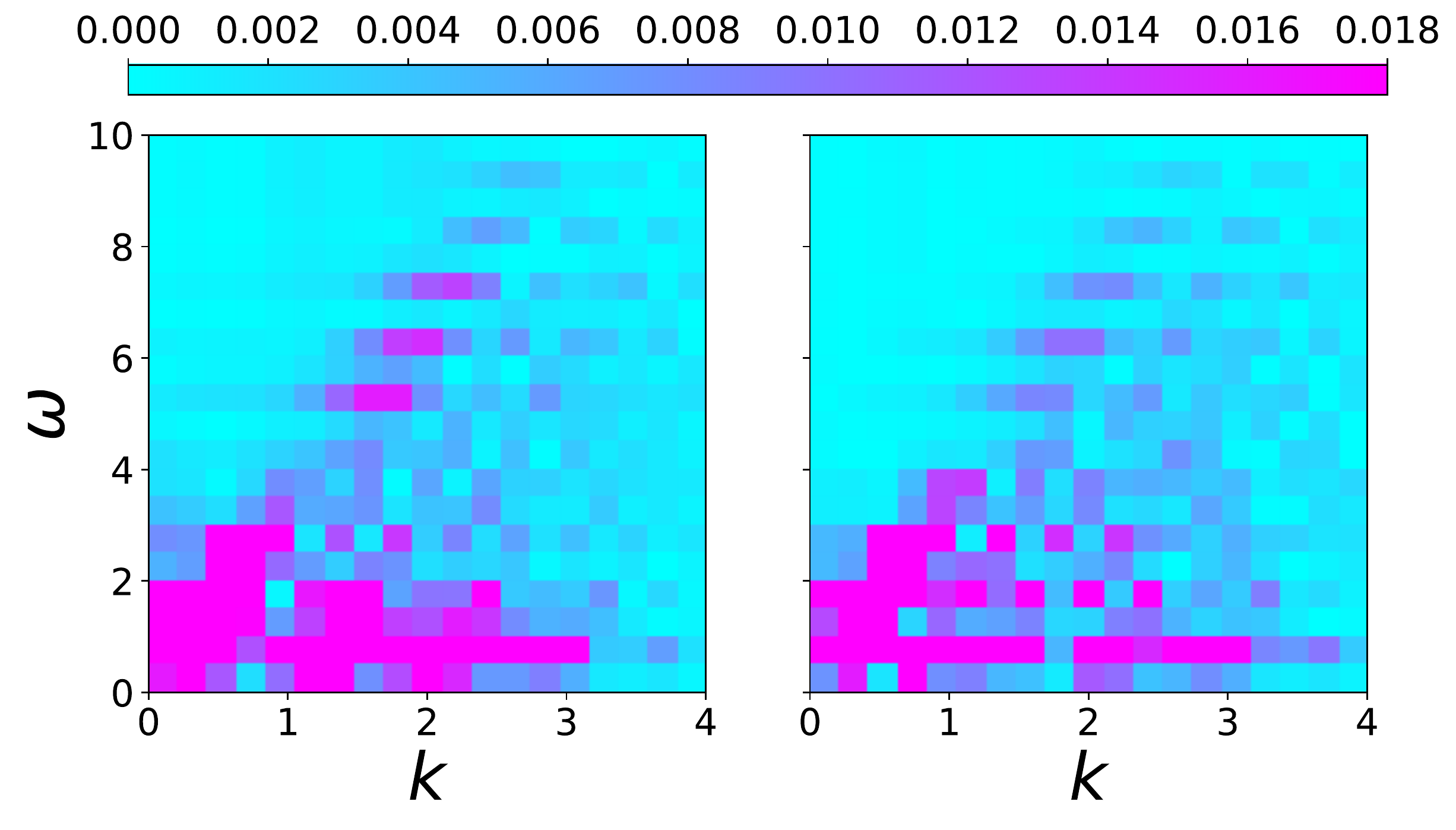}
	\caption{\label{dsf} (Color online) The dynamic structure factor (DSF) $S(k,\omega)$ for $\gamma=0.003$ (left) and $\gamma=1000$ (right). Horizontal axis is the momentum and the vertical axis is the energy transfer.}
\end{figure}

\paragraph{The structure factor.} The dynamical structure factor (DSF) is defined as
\begin{equation}
	S(k,\omega)=\int  e^{i(\omega t - kx)} \langle\rho(x,t)\rho(0,0)\rangle dx dt.
\end{equation}	
Previously this has been analytically calculated for the LL gas in the absence of an external trap in Caux {\em et al.}\cite{PhysRevA.74.031605}, and the static structure factor $S(k)=\int \frac{d\omega}{2\pi}S(k,\omega)$ has been obtained using quantum Monte Carlo techniques \cite{PhysRevA.68.031602}. Our numerical result for the DSF at interaction strengths $\gamma= 0.003$ and $1000$ is presented in Fig. \ref{dsf}. Here, the $\omega$ axis corresponds to the energy transfer and $k$ corresponds to momentum. According to the results, for weak interactions most of spectral weight of $S(k,\omega)$ is found in the vicinity of a type-I excitation. For strongly interacting bosons, the spectrum is marginally broadened. On the other hand, being far away from the thermodynamic limit ($N=3$) and due to the low resolution, a rigorous comparison with Boguliubov theory would be greatly ambitious.

\begin{figure*}[t]
	\begin{minipage}[]{0.48\linewidth}
		\includegraphics[width=\linewidth]{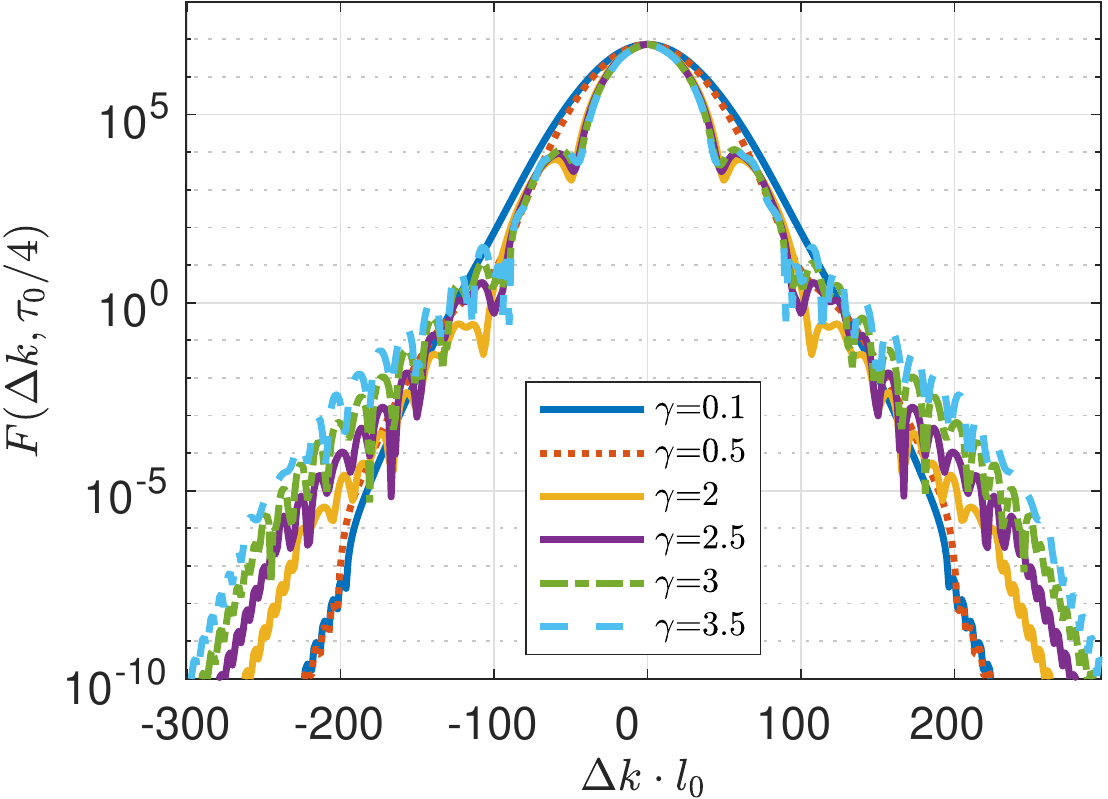}
	\end{minipage}
	\hfill
	\begin{minipage}[]{0.48\linewidth}
		\includegraphics[width=\linewidth]{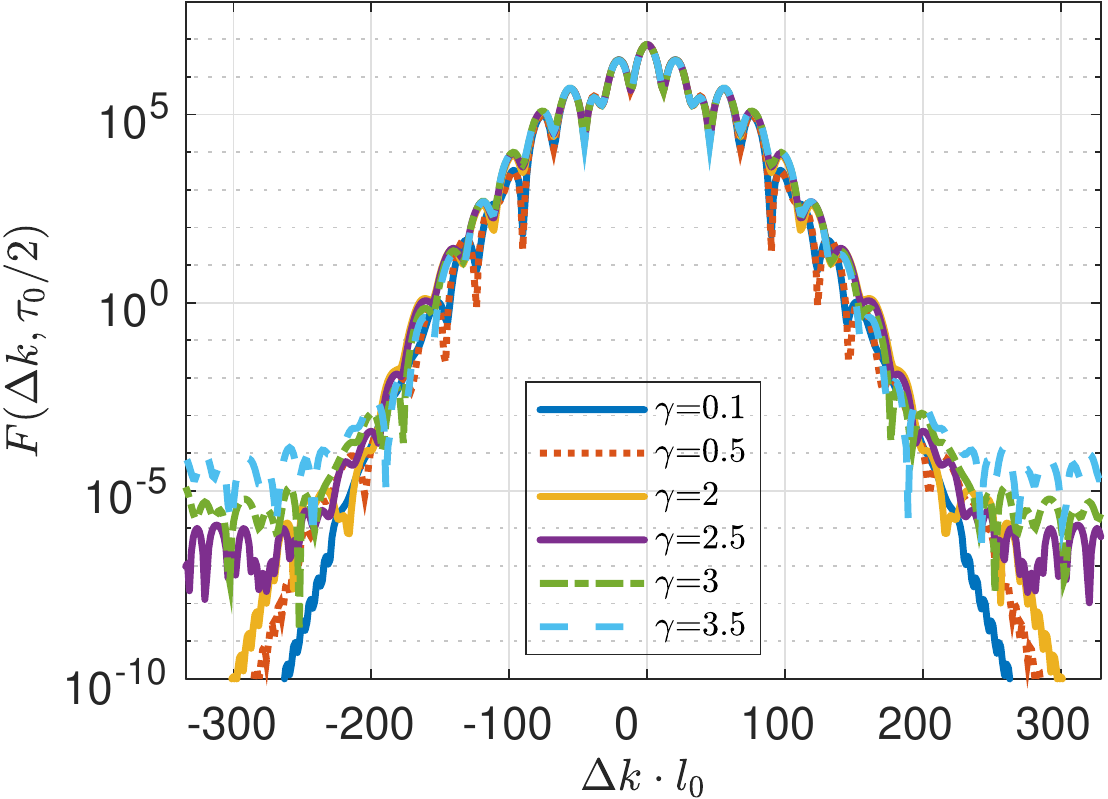}
	\end{minipage}
	\caption{\label{sf} (Color online) The plot of structure factor at the moment of three-body collision ($t=\tau_0/4$) in (a), and  at the moment when the oscillating particle is furthest from the equilibrium position ($t=\tau_0/2$) in (b). The high-momentum tail of $F(\Delta k,t)$ in (b) illustrate the dependence of momentum transfer on the interaction strength $\gamma$, and can be used for identification of the Tonks-Girardeau regime in experiments.}
\end{figure*}

From an experimental point of view, the dynamics of the system can be characterized by Bragg spectroscopy \cite{Sette1550}. One can treat the bosonic system as a perturbative potential that transfers a momentum of $\hbar \Delta k$ to the incoming light. Then the corresponding scattering amplitude can be quantified by the inelastic, time dependent structure factor \cite{1742-5468-2011-12-P12010}
\begin{equation}\label{structure}
	f(\Delta k,t)\propto\int e^{-i x \Delta k} \rho(x,t) d x,
\end{equation}
and it can be probed by light scattering experiments. The time evolution of the norm of the structure factor $F(\Delta k,t)=|f(\Delta k,t)|^2$ is given in the right panel of Fig. \ref{realspace}.b, which brings us to the third and the final main result presented in this Letter. It is found that the profile of the structure factor retains similar characteristics for different values of $\gamma$. Nonetheless, it is reported that for weaker interactions, the weight of $F(\Delta k,t)$ shifts towards smaller momentum components. This qualitative remark is reflected in the right panel of Fig. \ref{dsf}. It is seen that the most profound contrast between $\gamma=0.3$ and $\gamma=3$ cases occurs at times equal to odd multiples of $\tau_0/4$, corresponding to an overlap of the wave-packets. $\mathcal{T}(t)$ is related to the spread of the scattering amplitude at $t=\tau_0/4$. This suggests that, despite the indistinguishability of the particles, it is possible to acquire insight into a collision event of bosons through Bragg spectroscopy.

To obtain a quantitative account of the collision dynamics, the structure factor can be investigated for different values of $\gamma$ by focusing on $F(\Delta k,t)$ at specific time slices. In Fig. \ref{sf}, $F(\Delta k,t)$ is plotted (a) at $t=\tau_0/4$, i.e. at the moment of three-body collision, and (b) at $t=\tau_0/2$, i.e. when the initially displaced particle is furthest from the equilibrium position of the external trap. Here, for both cases, the basic observation is that high momentum transfer rates monotonically increase with increasing $\gamma$. The tails of the structure factor escalates with increasing $\gamma$ in accordance with Caux {\em et al.}\cite{1742-5468-2007-01-P01008} and Minguzzi {\em et al.}\cite{MINGUZZI2002222}. On the other hand, given that the infrared behavior is usually obscured by the harmonic trap \cite{tg_realisation}, it is also essential to note that these results also shed light into the previously concealed intermediate window of momenta.

\paragraph{Conclusion.} To summarize, the evolution of the reduced real-space density and the one-dimensional structure factor are computed for specific finite values of the two-body interaction strength $c$. The dependence of the motion period and the transmission amplitude of colliding bosons on the numerical interaction parameter $\gamma$ is demonstrated. It is suggested that the period of motion deviates from the characteristic period of the harmonic trap at a maximum about 6\%. From the empirical perspective, this constitutes a method for establishing the link between $\gamma$ and the experimental parameters.

\bibliography{bib}

\end{document}